\documentclass[12pt]{article}
\usepackage[left=2.5cm,top=2.50cm,right=2.5cm,bottom=2.50cm]{geometry}
\usepackage{mathrsfs}
\usepackage{amsmath,amssymb,latexsym,color,cancel,graphicx,bbm,colortbl}
\usepackage[english]{babel}
\usepackage[latin1]{inputenc}
\usepackage{ragged2e}
\usepackage{cite}
\begin{document}
\date{}

\title{Landau levels for the $(2+1)$ Dunkl-Klein-Gordon oscillator}
\author{R. D. Mota$^{a}$, D. Ojeda-Guill\'en$^{b}$\footnote{{\it E-mail address:} dojedag@ipn.mx},\\ M. Salazar-Ram{\'i}rez$^{b}$ and V. D. Granados$^{c}$} \maketitle

\begin{minipage}{0.9\textwidth}
\small $^{a}$ Escuela Superior de Ingenier{\'i}a Mec\'anica y El\'ectrica, Unidad Culhuac\'an,
Instituto Polit\'ecnico Nacional, Av. Santa Ana No. 1000, Col. San
Francisco Culhuac\'an, Del. Coyoac\'an, C.P. 04430, Ciudad de M\'exico, Mexico.\\

\small $^{b}$ Escuela Superior de C\'omputo, Instituto Polit\'ecnico Nacional,
Av. Juan de Dios B\'atiz esq. Av. Miguel Oth\'on de Mendiz\'abal, Col. Lindavista,
Del. Gustavo A. Madero, C.P. 07738, Ciudad de M\'exico, Mexico.\\

\small $^{c}$ Escuela Superior de F{\'i}sica y Matem\'aticas,
Instituto Polit\'ecnico Nacional, Ed. 9, Unidad Profesional Adolfo L\'opez Mateos, Del. Gustavo A. Madero, C.P. 07738, Ciudad de M\'exico, Mexico.\\

\end{minipage}

\begin{abstract}

In this paper we study the  $(2+1)$-dimensional Klein-Gordon oscillator coupled to an external magnetic field, in which we change the standard partial derivatives for the Dunkl derivatives. We find the energy spectrum (Landau levels) in an algebraic way, by introducing three operators that close the $su(1,1)$ Lie algebra and from the theory of unitary representations. Also we find the energy spectrum and the eigenfunctions analytically, and we show that both solutions are consistent. Finally, we demonstrate that when the magnetic field vanishes or when the parameters of the Dunkl derivatives are set zero, our results are adequately reduced to those reported in the literature.

\end{abstract}

PACS: 02.30.Ik, 02.30.Jr, 03.65.Ge, 03.65.Pm\\
Keywords: Dunkl derivative, exact solutions, Klein-Gordon oscillator, Landau levels

\section{Introduction}

Reflection operators were introduced in quantum mechanics by Wigner to generalize the quantization rules \cite{wigner}. However, Yang was the first to apply these operators to the one-dimensional harmonic oscillator \cite{yang}. These reflection operators are very useful to study the quantum Calogero and Calogero-Sutherland-Moser models and their integrability \cite{br,ply,hikami,kakei,lapon}.

Dunkl reintroduced the reflection operators for the study of functions of several variables with discrete symmetry groups. The now known as Dunkl operators are operators which consist of combinations of a differential part and a discrete part \cite{dunkl1}. One of the main applications of the Dunkl operators has been the study of the Laplace operator in  $\bf  R^n$. The Dunkl-Laplace operator has been shown to consist of two parts, where one part is equal to classical Laplacian and is invariant with respect to the orthogonal group, and the other part depends on the reflection operators and is invariant under the reflection group \cite{dunkl1,dunkl2}.

The Dunkl operators are closely related to the Bannai-Ito and the Dunkl-Schwinger algebras. Dunkl operators have been essentially applied to solve central field problems in two and three dimensions in terms of the Jacobi, Legendre, Hermite and $-1$ orthogonal polynomials  \cite{GEN1,GEN2,GEN3,GEN4,nos1,nos2}. It has also been applied to the study of new integrable systems \cite{GEN1,GEN4}.
Recently, the Schr\"odinger equation for the Dunkl-Coulomb problem in $3D$ has been solved and its superintegrability and dynamical symmetry have been studied \cite{sami1,sami2}. In the relativistic regime, we studied the problem of the Dirac-Dunkl oscillator in two dimensions in Ref. \cite{nos3}, and the Dunkl-Klein-Gordon equation for the Coulomb potential and the Klein-Gordon oscillator has been algebraically and analytically solved in Ref. \cite{arxiv}. Currently, this derivative is being widely studied in many branches of physics, from classical mechanics, to relativistic quantum mechanics, through electromagnetism, as can be seen in the Refs. \cite{Chung1,Chung2,Chung3,Chung4}.

In this paper we study the $2D$ Dunkl-Klein-Gordon oscillator coupled to a uniform magnetic field. We show in two different ways that this problem can be solved exactly. In the first way, we find the energy spectrum from the theory of unitary representations by introducing a set of operators which close the $su(1,1)$ Lie algebra. In the second way, we solve the Dunkl-Klein-Gordon equation analytically and find the energy spectrum and the eigenfunctions.

This work is organized as follows. In Section 2, we obtain the Dunkl-Klein-Gordon equation in $2D$ for the Klein-Gordon oscillator coupled to an external magnetic field. We show that the generalization with the Dunkl derivative of the $z$-component of the angular momentum is the one that allows the separation of variables for the Dunkl-Klein-Gordon equation. Then, we give the $z$-component of the angular eigenfunctions and the radial equations for each of the two cases in which the Dunkl angular momentum operator can be studied. In Section 3, we introduce a set of operators which close the $su(1,1)$ Lie algebra to obtain the energy spectrum of this problem from an algebraic point of view. In Section 4, the energy spectrum and eigenfunctions of the Dunkl-Klein-Gordon equation for Klein-Gordon oscillator coupled to an external magnetic field are found analytically. Finally, in Section $5$, we give some concluding remarks.

\section{$2D$ Dunkl-Klein-Gordon oscillator coupled to an external magnetic field}

The problem of a charged particle in a uniform magnetic field has been widely studied in many branches of physics, as classical mechanics, condensed matter physics, quantum optics and relativistic quantum mechanics, among others. The energy spectrum of this problem is known as the Landau levels. The interaction of an electron with the uniform magnetic field is described by means of electromagnetical potentials. The interesting thing about this problem is that different gauges give raise to the same electromagnetic field \cite{Capri}.

The standard Klein-Gordon oscillator equation for stationary states in $2D$  is \cite{bruce,mex,roa,boumali,bras}
\begin{equation}
(E^2-m^2c^4)\Psi_{O}=c^2\left({\bf P}+im\omega\rho \hat \rho\right)\cdot \left({\bf P}-im\omega\rho \hat \rho\right)\Psi_{O},
\end{equation}
where $\rho=\sqrt{x^2+y^2}$ and $\hat \rho$ is a unitary radial vector. To study the Landau levels, the presence of a uniform magnetic field $\bf B$ is  incorporated by minimal coupling $ {{\bf P}} \rightarrow {{\bf P}}-\frac{e}{c}{\bf A}$. We set the vector potential in the symmetric gauge
${\bf A }= (-\frac{By}{2}, \frac{Bx}{2})$ and $e=-|e|$, being $|e|$ the electron charge. Thus, the Klein-Gordon equation takes the form
\begin{equation}
\left({\bf P}^2+\frac{2|e|}{c}{\bf A}\cdot {\bf P} +im\omega(xP_1- P_1x+yP_2-P_2y)+m^2\Omega^2\rho^2\right)\Psi_{O}=\epsilon'\Psi_{O},\label{full}
\end{equation}
where the definitions
\begin{equation}
\epsilon'\equiv \frac{E^2-m^2c^4}{c^2},\hspace{10ex}m^2\Omega^2\equiv m^2\omega^2+\frac{|e|^2B^2}{4c^2},\label{def}
\end{equation}
were introduced. From Eqn. (\ref{full}) we observe that second term vanishes when we eliminate the magnetic field. On the other hand, the last of Eqn. (\ref{def}) implies that the magnetic field is responsible for an increase in the frequency of the oscillator. When the magnetic field vanishes, then $\Omega=\omega$.

We do not simplify more Eqn. (\ref{full}), since we are interested in incorporating the Dunkl derivative.
If we change the standard partial derivatives $\frac{\partial}{\partial x}$ and $\frac{\partial}{\partial y}$ by the Dunkl derivatives
\begin{equation}
D_1\equiv\frac{\partial}{\partial x}+\frac{\mu_1}{x}(1-R_1), \quad\quad D_2\equiv\frac{\partial}{\partial y}+\frac{\mu_2}{y}(1-R_2),
\end{equation}
we obtain the Dunkl-Klein-Gordon ($DKG$) equation. In this definition, the constants $\mu_1,\mu_2$ satisfy $\mu_1>0$ and $\mu_2>0$ \cite{GEN4}, and $R_1,R_2$ are the reflection operators with respect to the $x-$ and $y-$ coordinates, it is to say, $R_1f(x,y)=f(-x,y)$ and $R_2f(x,y)=f(x,-y)$. Therefore, ${\bf P}$ changes to $-i\hbar(D_1,D_2)$, and ${\bf P}^2=-\hbar^2 \nabla^2$ changes to
${\bf P}^2=-\hbar^2\left(D_1^2+D_2^2\right)\equiv-\hbar^2 \nabla^2_D$, where  $ \nabla^2_D$ is known as the Dunkl Laplacian. Hence, the $DKG$ equation for a particle in a uniform magnetic field is
\begin{equation}
\left(-\hbar^2{\nabla }_D^2-i\frac{\hbar |e|B}{c}(xD_2-yD_1)+m^2\Omega^2\rho^2 +\hbar m\omega(xD_1- D_1x+yD_2-D_2y)\right)\Psi_{O}=\epsilon'\Psi_{O}.\label{ec1}
\end{equation}

The action of the reflection operator $R_i$ on a two variables function $f(x,y)$ implies
\begin{equation}
R_1D_1=-D_1R_1,\hspace{3ex} R_1^2=1, \hspace{3ex}\frac{\partial}{\partial x}R_{1}=-R_1\frac{\partial}{\partial x}, \hspace{3ex}R_1x=-xR_1,\label{pro1}
\end{equation}
and similar expressions for the $y-$ coordinate. Also, the following equalities involving the operators $R_i$ and $D_i$ can be proved
\begin{equation}
R_1R_2=R_2R_1,\hspace{2ex}[D_1,D_2]=0,\hspace{2ex}[x_i,D_j]=\delta_{ij}+2\mu_{\delta{ij}}R_{\delta{ij}}\hspace{1ex}\hbox{(no sum over $i$ and $j$)}.\label{pro2}
\end{equation}
Thus, using the last property, we can write equation (\ref{ec1}) as
\begin{equation}
H_O\Psi_{O}\equiv \left(-\hbar^2{\nabla }_D^2-i\frac{\hbar |e|B}{c}(xD_2-yD_1)+m^2\Omega^2\rho^2 +2\hbar m\omega(1+\mu_1R_1+\mu_2R_2)\right)\Psi_{O}=\epsilon'\Psi_{O},\label{DKGOO}
\end{equation}
where we have defined the Hamiltonian $H_O$.

Now, we introduce the Dunkl angular momentum ${\mathcal J}=i(xD_2-yD_1)$, which can be used to show the following results
\begin{eqnarray}
&&\left[xD_2,\nabla_D^2 \right]=2D_2D_1,\label{res1}\\
&&\left[yD_1,\nabla_D^2\right]=2D_1D_2,\label{res2}\\
&&\left[\frac{\mu_i}{x_i}\left(1-R_i\right),F(\rho)\right]=0, \hspace{2ex}i=1,2,\label{res3}\\
&&\left[\left(x\frac{\partial}{\partial y}-y\frac{\partial}{\partial x}\right),F(\rho)\right]=\left[\frac{\partial }{\partial \phi},F(\rho)\right]=0,\label{res4}
\end{eqnarray}
where $f(\rho)$ is an arbitrary function with partial derivative. In the last two equalities we have used the polar coordinates $\rho=\sqrt{x^2+y^2}$, $\tan{\phi}=\frac{y}{x}$. From these commutation relations, we immediately show that the operator $\mathcal J$  is a constant of motion of the Hamilton operator $H_O$
\begin{equation}
[{\mathcal J}, H_O]=0.
\end{equation}
As it will be shown below, this fact will allow us to solve the $DKG$ equation (\ref{DKGOO}) by using separation of variables on the $DKG$ wave function.

Explicitly, the Dunkl Laplacian in cartesian coordinates takes the form
\begin{eqnarray}
&&\nabla_D^2= D_1^2+D_2^2\\
&&\hspace{4ex}=\frac{\partial^2}{\partial x^2}+\frac{\partial^2}{\partial y^2}+2\frac{\mu_1}{x}\frac{\partial}{\partial x}+2\frac{\mu_2}{y}\frac{\partial}{\partial y}+\frac{\mu_1}{x^2}(1-R_1)+\frac{\mu_2}{y^2}(1-R_2),
\end{eqnarray}
or in  polar coordinates it is written as
\begin{equation}
\nabla_D^2= \frac{\partial^2}{\partial \rho^2}+\frac{1+2\mu_1+2\mu_2}{\rho}\frac{\partial}{\partial \rho}-\frac{2}{\rho^2}B_\phi, \label{laplapol}
\end{equation}
where the operator $B_\phi$ is given by
\begin{equation}
B_\phi\equiv-\frac{1}{2}\frac{\partial^2}{\partial \phi^2}+\left(\mu_1\tan{\phi}-\mu_2\cot{\phi}\right)\frac{\partial}{\partial \phi}
+\frac{\mu_1 (1-R_1)}{2\cos^2{\phi}}+\frac{\mu_2 (1-R_2)}{2\sin^2{\phi}}.
\end{equation}

In polar coordinates the operator $\mathcal J$ takes the form
\begin{equation}
\mathcal{J}=i(\partial_\phi+\mu_2\cot\phi(1-R_2)-\mu_1\tan\phi(1-R_1)).\label{j}
\end{equation}
Some direct calculations leads us to show that the square of this operator can be written as
\begin{equation}
\mathcal{J}^2=2B_\phi+2\mu_1\mu_2(1-R_1R_2).\label{jcuad}
\end{equation}
Substituting the results of equations (\ref{laplapol}), (\ref{j}) and (\ref{jcuad}), into equation (\ref{DKGOO}), we obtain
\begin{equation}
\left(-\frac{\partial^2}{\partial \rho^2}-\frac{1+2\mu_1+2\mu_2}{\rho}\frac{\partial}{\partial \rho}+\frac{{\mathcal J}^2-2\mu_1\mu_2(1-R_1R_2)}{\rho^2}-\frac{|e|B}{\hbar c}\mathcal J +\frac{m^2\Omega^2}{\hbar^2}\rho^2\right)\Psi_O=\tilde{\epsilon} \Psi_O\label{rado},
\end{equation}
where
\begin{equation}
\tilde{\epsilon} \equiv \left(\frac{E^2-m^2c^4}{\hbar^2c^2}\right)-\frac{2m\omega}{\hbar}(1+\mu_1R_1+\mu_2R_2).
\end{equation}

Considering that $[R_1,H_O]=[R_2,H_O]=0$ and $[R_1R_2,H_O]$, the eigenvalues and eigenfunctions of the operator $\mathcal J$ have been constructed in Ref. \cite{GEN1}. Here, we shall give a summary of the main results.
Since the operator $R_1R_2$ commutes with the operator $\mathcal{J}$, its eigenvalues and eigenvectors are search in the form
\begin{equation}
\mathcal{J}F_\epsilon=\lambda_\epsilon F_\epsilon,\label{angular}
\end{equation}
being $\epsilon\equiv s_1s_2=\pm 1$, and $s_1$, $s_2$ the eigenvalues of the reflection operators $R_1$ and $R_2$, respectively. There are two cases in which the eigenfunctions and eigenvalues of the operator $\mathcal J$ are classified.\\

Case A) If $R_1=R_2$, then $\epsilon=1$. The solutions of equation (\ref{angular}) are given by
\begin{eqnarray}
F_+=\Phi_\ell^{++}(\phi)\pm i\Phi_\ell^{--}(\phi),\hspace{.5ex}\\
\lambda_+=\pm2\sqrt{\ell(\ell+\mu_1+\mu_2)}
\end{eqnarray}
where $\ell\in {\mathbb{N}}$. The functions $\Phi_\ell^{++}$ and $\Phi_\ell^{--}$ explicitly are
\begin{eqnarray}
\Phi_\ell^{++}(x)=\sqrt{\frac{(2\ell+\mu_1+\mu_2)\Gamma{(\ell+\mu_1+\mu_2)\ell !}}{2\Gamma{(\ell+\mu_1+1/2)}\Gamma{(\ell+\mu_2+1/2)}}}P_\ell^{(\mu_1-1/2,\mu_2-1/2)}(x),\hspace{18ex}\label{masmas}\\
\Phi_\ell^{--}(x)=\sqrt{\frac{(2\ell+\mu_1+\mu_2)\Gamma{(\ell+\mu_1+\mu_2+1)(\ell-1)!}}{2\Gamma{(\ell+\mu_1+1/2)}\Gamma{(\ell+\mu_2+1/2)}}}\sin\phi\cos\phi P_{\ell-1}^{(\mu_1+1/2,\mu_2+1/2)}(x),\label{menmen}
\end{eqnarray}
where $P_\ell^{(\alpha,\beta)} (x)$ are the classical Jacobi polynomials and $x=-\cos2\phi$. It is understood that $P_{-1}^{(\alpha,\beta)}(x)=0$ and hence that $\Phi_0^{--}=0$.\\

Case B) For $R_1=-R_2$, $\epsilon=-1$, it has been shown that
\begin{eqnarray}
F_-=\Phi_\ell^{-+}(\phi)\mp i\Phi_\ell^{+-}(\phi),\hspace{3ex}\\
\lambda_-=\pm2\sqrt{(\ell+\mu_1)(\ell+\mu_2)},
\end{eqnarray}
where $\ell\in \{\frac{1}{2},\frac{3}{2},...\}$. The eigenfunctions $\Phi_\ell^{-+}$ and $\Phi_\ell^{+-}$ are given by
\begin{eqnarray}
\Phi_\ell^{-+}(x)=\sqrt{\frac{(2\ell+\mu_1+\mu_2)\Gamma{(\ell+\mu_1+\mu_2+1/2)(\ell-1/2)!}}{2\Gamma{(\ell+\mu_1+1)}\Gamma{(\ell+\mu_2)}}}\cos\phi P_{\ell-1/2}^{(\mu_1+1/2,\mu_2-1/2)}(x),\label{menmas}\\
\Phi_\ell^{+-}(x)=\sqrt{\frac{(2\ell+\mu_1+\mu_2)\Gamma{(\ell+\mu_1+\mu_2+1/2)(\ell-1/2)!}}{2\Gamma{(\ell+\mu_1)}\Gamma{(\ell+\mu_2+1)}}}\sin\phi P_{\ell-1/2}^{(\mu_1-1/2,\mu_2+1/2)}(x).\hspace{.5ex}\label{masmen}
\end{eqnarray}

In what follows we define $r=\sqrt{\frac{m \Omega}{\hbar}}\rho$. According to these results, we divide the solutions of Eqn. (\ref{rado}) in two cases. I) If $(R_1,R_2)$ take the values $(s_1,s_2)=(1,1)$ or $(s_1,s_2)=(-1,-1)$, and  II) If $(R_1,R_2)$ take the values $(s_1,s_2)=(-1,1)$ or $(s_1,s_2)=(1,-1)$.  \\

Case I) In this case, the eigenvalue of the operator $\mathcal J$ is equal to $\lambda_+$. Thus, from Eqn. (\ref{rado}), the radial equation that we must solve is\\
\begin{equation}
\left(-\frac{d^2}{d r^2}-\frac{1+2\mu_1+2\mu_2}{r}\frac{d}{d r}+\frac{\lambda_+^2}{r^2}+r^2\right)R(r)=\tilde{\mathcal E}R(r)\label{rad1},
\end{equation}
with
\begin{equation}
\tilde{\mathcal E} \equiv \left(\frac{E^2-m^2c^4}{\hbar m\Omega c^2}\right)-2\left(\frac{\omega}{\Omega}\right)(1\pm\mu_1\pm\mu_2)+\frac{|e|B}{\hbar m \Omega  c}\lambda_+.\label{espec1}
\end{equation}
The upper signs correspond to $(s_1,s_2)=(1,1)$ and the lower signs correspond to $(s_1,s_2)=(-1,-1)$.

Case II) The eigenvalue of the operator $\mathcal J$ is $\lambda_-$. Hence, the radial equation for this case results to be
\begin{equation}
\left(-\frac{d^2}{d r^2}-\frac{1+2\mu_1+2\mu_2}{r}\frac{d}{d r}+\frac{\lambda_-^2-4\mu_1\mu_2}{r^2}+r^2\right)R(r)=\tilde{\mathcal E}'R(r)\label{rad2},
\end{equation}
with
\begin{equation}
\tilde{\mathcal E}' \equiv \left(\frac{E^2-m^2c^4}{\hbar m\Omega c^2}\right)-2\left( \frac{\omega}{\Omega}\right)(1\mp\mu_1\pm\mu_2)+\frac{|e|B}{\hbar m \Omega  c}\lambda_-.\label{espec2}
\end{equation}
In this Eqn. the upper signs correspond to $(s_1,s_2)=(-1,1)$ and the lower signs correspond to $(s_1,s_2)=(1,-1)$.
In the next Section we will solve the radial Eqns. (\ref{rad1}) and (\ref{rad2}) using an algebraic approach.

\section{Algebraic approach for the Landau levels for the $DKG$-oscillator}

The three generators $K_\pm=K_1\pm i K_2$, and $K_{0}$ , which satisfy the commutation relations
\begin{eqnarray}
[K_{0},K_{\pm}]=\pm K_{\pm},\quad\quad [K_{-},K_{+}]=2K_{0},\label{com}
\end{eqnarray}
define the $su(1,1)$ Lie algebra \cite{vourdas}.
The action of these operators on the Sturmian basis $\{|k,n\rangle, n=0,1,2,...\}$ is given by
\begin{eqnarray}
&&K_{+}|k,n\rangle=\sqrt{(n+1)(2k+n)}|k,n+1\rangle,\label{k+n}\\
&&K_{-}|k,n\rangle=\sqrt{n(2k+n-1)}|k,n-1\rangle,\label{k-n}\\
&&K_{0}|k,n\rangle=(k+n)|k,n\rangle,\label{k0n}
\end{eqnarray}
where $|k,0\rangle$ represent the lowest normalized state. We must emphasize that the equations (\ref{k+n})-(\ref{cas}) define the unitary irreducible representations of the $su(1,1)$ Lie algebra. Thus, the Bargmann's number $k$ completely determines a representation of the $su(1,1)$ Lie algebra. In this work, we will consider the discrete series only, in which $k>0$. The Casimir operator for any irreducible representation of the $su(1,1)$ Lie algebra is
\begin{equation}
K^{2}=K_0^2-K_1^2-K_2^2=-K_{+}K_{-}+K_{0}(K_{0}-1)=k(k-1).\label{cas}
\end{equation}

To solve equations (\ref{rad1}) and (\ref{rad2}) by algebraic methods, we introduce a set of operators which close the $su(1,1)$ Lie algebra. This set of operators are analogous to those we used to solve the Shr\"odinger equation for the $2D$ harmonic oscillator \cite{nos1}.\\

Thus, for the Eqn. (\ref{rad1}) of case I) we find the $su(1,1)$ generators
\begin{eqnarray}
&&O_0=\frac{1}{4}\left(-\frac{d^2}{dr^2}-\frac{1+2\mu_1+2\mu_2}{r}\frac{d}{dr}+\frac{\lambda_+^2}{r^2}+r^2\right),\\\label{o1}
&&O_+=\frac{1}{2}\left(-r\frac{d}{dr}+r^2 -(1+\mu_1+\mu_2)-2O_0\right),\\
&&O_-=\frac{1}{2}\left(r\frac{d}{dr}+r^2 +(1+\mu_1+\mu_2)-2O_0\right).\label{o3}
\end{eqnarray}
 The Casimir operator for this algebra is obtained by direct computation, and is given by
\begin{equation}
O^2=\frac{\lambda_+^2+(\mu_1+\mu_2)^2-1}{4}.
\end{equation}
According to the $su(1,1)$ representation theory, this value must be equal to $k(k-1)$. From this fact, we obtain that for the discrete series $k>0$,
\begin{equation}
k=\frac{1}{2}\left(1+\sqrt{\lambda_+^2+(\mu_1+\mu_2)^2}\right)=\ell+\frac{1+\mu_1+\mu_2}{2},
\end{equation}
where we have used that $\lambda_+^2=4\ell(\ell+\mu_1+\mu_2)$. By writing the left hand side of Eqn. (\ref{rad1}) in terms of the $O_0$ operator, and using Eqn. (\ref{k0n}), we have
\begin{equation}
O_0R(r)=(n+k)R(r)=\frac{1}{4}\tilde{\mathcal{E}} R(r).
\end{equation}
From the second equality and the definition of $\tilde{\mathcal E}$ (equation (\ref{espec1})), we get that the energy spectrum is given by
\begin{equation}
E=\pm mc^2\left(1+\frac{4\hbar \Omega}{mc^2}\left(n+\ell +\frac{1}{2}\left(1+\frac{\omega}{\Omega}\right)+\frac{\mu_1}{2}\left(1\pm\frac{\omega}{\Omega}\right)+\frac{\mu_2}{2}\left(1\pm\frac{\omega}{\Omega}\right)-\frac{|e|B(\ell+\mu_1+\mu_2)}{\hbar m \Omega  c}\right)\right)^\frac{1}{2}.\label{espectro1}
\end{equation}
When the magnetic field vanishes $B=0$, $\omega=\Omega$, and this spectrum is in full agreement with the spectrum reported in Ref. \cite{arxiv} for the cases $(R_1,R_2)=(1,1)$ and $(R_1,R_2)=(-1,-1)$.\\

Analogously, for the Eqn. (\ref{rad2}) of case II) we set the $su(1,1)$ Lie algebra generators as
\begin{eqnarray}
&&{\mathcal O}_0=\frac{1}{4}\left(-\frac{d^2}{dr^2}-\frac{1+2\mu_1+2\mu_2}{r}\frac{d}{dr}+\frac{\lambda_-^2-4\mu_1\mu_2}{r^2}+r^2\right),\\\label{oo1}
&&{\mathcal O}_+=\frac{1}{2}\left(-r\frac{d}{dr}+r^2 -(1+\mu_1+\mu_2)-2{\mathcal O}_0\right),\\
&&{\mathcal O}_-=\frac{1}{2}\left(r\frac{d}{dr}+r^2 +(1+\mu_1+\mu_2)-2{\mathcal O}_0\right).\label{oo3}
\end{eqnarray}
Here, the Casimir operator results to be
\begin{equation}
O^2=\frac{\lambda_-^2-4\mu_1\mu_2+(\mu_1+\mu_2)^2-1}{4},
\end{equation}
and equating this value with $k(k-1)$, we find that the positive Bargmann's number is
\begin{equation}
k=\ell+\frac{1+\mu_1+\mu_2}{2}.
\end{equation}
Similarly to the previous case, using the unitary theory of representations of the $su(1,1)$ Lie algebra, we obtain $\frac{1}{4}\tilde{\mathcal E}'=n+1$.  Thus, by using Eqn. (\ref{espec2}) we obtain the energy spectrum
\begin{equation}
E=\pm mc^2\left(1+\frac{4\hbar \Omega}{mc^2}\left(n+\ell +\frac{1}{2}\left(1+\frac{\omega}{\Omega}\right)+\frac{\mu_1}{2}\left(1\mp\frac{\omega}{\Omega}\right)+\frac{\mu_2}{2}\left(1\pm\frac{\omega}{\Omega}\right)-\frac{|e|B(\ell+\mu_1)(\ell+\mu_2)}{\hbar m \Omega  c}\right)\right)^\frac{1}{2}.\label{espectro2}
\end{equation}
Again, when the magnetic field vanishes $B=0$, $\omega=\Omega$, and this spectrum is completely reduced to that found in the previous work \cite{arxiv} for the cases  $(R_1,R_2)=(-1,1)$ and $(R_1,R_2)=(1,-1)$.

\section{Analytical solution for the Landau levels of the  $DKG$-oscillator}

In this Section we shall solve the $DKG$ equation for the Landau levels of $DKG$ oscillator, Eqn. (\ref{rad1}) in an analytical form. To this end, we consider the differential equation
\begin{equation}
\left(-\frac{d^2}{d r^2}-\frac{\mathcal A}{r}\frac{d}{d r}+\frac{\mathcal B}{r^2}+r^2\right)R(r)={\mathcal E}R(r).\label{rado2}
\end{equation}
Setting $R(r)=r^{-\frac{1}{2}\mathcal A} G(r)$, and rearranging we obtain
\begin{equation}
\left(\frac{d^2}{d r^2}+{\mathcal E}- r^2+\frac{\frac{1}{2}{\mathcal A}-\frac{1}{4}{\mathcal A}^2-{\mathcal B}}{r^2}\right)G(r)=0.\label{nosotros}
\end{equation}
This equation has the same form of the differential equation
 \begin{equation}
u''+\left(4n+2\alpha+2-x^2+\frac{\frac{1}{4}-\alpha^2}{x^2}\right)u=0,\label{levedev}
\end{equation}
which has as solution the functions\cite{LEB}
\begin{equation}
u(x)=C_0e^{-\frac{x^2}{2}}x^{\alpha+\frac{1}{2}}L_n^\alpha(x^2)\hspace{5ex} n=0,1,2,...\label{ansol}
\end{equation}
where $C_0$ is a normalization constant. Thus, the comparison between equations (\ref{nosotros}) and (\ref{levedev}) leads to
\begin{equation}
\alpha=\frac{1}{2}\sqrt{({\mathcal A}-1)^2+4{\mathcal B}},\hspace{7ex} {\mathcal{E}}=4n+2+\sqrt{({\mathcal A}-1)^2+4{\mathcal B}}.\label{uno}
\end{equation}
The equations (\ref{rad1}) and (\ref{rado2}) coincide with the identifications
\begin{equation}
 \tilde {\mathcal E}={\mathcal E}, \hspace{10ex}{\mathcal B}=\lambda_+^2=4\ell(\ell+\mu_1+\mu_2),\hspace{10ex} {\mathcal A}=1+2\mu_1+2\mu_2.\label{dos}
\end{equation}
With these particular values it results that
\begin{equation}
\alpha=2\ell+\mu_1+\mu_2.\label{alfa1}
\end{equation}
Using these expressions and Eqns. (\ref{espec1}), (\ref{uno}) and (\ref{dos}), we immediately show that the resulting energy spectrum is the same as that given by the equation (\ref{espectro1}),
\begin{equation}
E=\pm mc^2\left(1+\frac{4\hbar \Omega}{mc^2}\left(n+\ell +\frac{1}{2}\left(1+\frac{\omega}{\Omega}\right)+\frac{\mu_1}{2}\left(1\pm\frac{\omega}{\Omega}\right)+\frac{\mu_2}{2}\left(1\pm\frac{\omega}{\Omega}\right)-\frac{|e|B(\ell+\mu_1+\mu_2)}{\hbar m \Omega  c}\right)\right)^\frac{1}{2}.\label{espectro11}
\end{equation}
The comparison between Eqns. (\ref{rad2}) and (\ref{rado2}), leads to the identifications
 \begin{equation}
 \tilde {\mathcal E}'={\mathcal E}, \hspace{10ex}{\mathcal B}=\lambda_-^2-4\mu_1\mu_2=4\ell(\ell+\mu_1+\mu_2),\hspace{10ex} {\mathcal A}=1+2\mu_1+2\mu_2.\label{dosprima}
\end{equation}
Hence, these values imply that
\begin{equation}
\alpha=2\ell+\mu_1+\mu_2. \label{alfa2}
\end{equation}
A simple calculation leads us to show that the energy spectrum obtained in this way is equal as that we found in Eqn. (\ref{espectro2})
\begin{equation}
E=\pm mc^2\left(1+\frac{4\hbar \Omega}{mc^2}\left(n+\ell +\frac{1}{2}\left(1+\frac{\omega}{\Omega}\right)+\frac{\mu_1}{2}\left(1\mp\frac{\omega}{\Omega}\right)+\frac{\mu_2}{2}\left(1\pm\frac{\omega}{\Omega}\right)-\frac{|e|B(\ell+\mu_1)(\ell+\mu_2)}{\hbar m \Omega  c}\right)\right)^\frac{1}{2}.\label{espectro22}
\end{equation}
We notice that Eqns. (\ref{dos}), (\ref{alfa1}), (\ref{dosprima}) and  (\ref{alfa2}) imply that the values of $\mathcal A$, $\mathcal B$ and $\alpha$  for the expressions (\ref{rad1}) and (\ref{rad2}) are equal. Thus, the radial eigenfunctions given by Eqn. (\ref{ansol}) are the same for the cases I) and II). However, since the right hand side of Eqns. (\ref{rad1}) and (\ref{rad2}) are different, it has as a consequence that the energy spectrum of Eqns. (\ref{espectro11}) and (\ref{espectro22}) are different.

For this problem the normalization constant $C_0$ can be determined by using $\alpha=2\ell+\mu_1+\mu_2$ and the orthogonality of the Laguerre polynomials \cite{LEB}
\begin{equation}
\int_0^{\infty}e^{-x}x^{\alpha}\left[L_{n}^{\alpha}(x)\right]^2dx=\frac{\Gamma(n+\alpha+1)}{n!}.\label{norm}
\end{equation}
With these results the normalization constant $C_0$ results to be
\begin{equation}
C_0=\sqrt{\frac{2n!}{\Gamma(n+2\ell+\mu_1+\mu_2+1)}}.
\end{equation}
Consequently, the eigenfunctions of the $DKG$ equation for the Klein-Gordon oscillator coupled to
an external magnetic field are explicitly given by
\begin{equation}
R_{n\ell}(r)=\sqrt{\frac{2n!}{\Gamma(n+2\ell+\mu_1+\mu_2+1)}}e^{-\frac{r^2}{2}}r^{2\ell}L_n^{2\ell+\mu_1+\mu_2}(r^2).
\end{equation}
In this way we have computed the energy spectrum and eigenfunctions of the Landau levels for the $DKG$ oscillator. We have shown that the analytical and algebraic methods used in this work are in complete agreement.

\section{Concluding Remarks}

We have exactly solved the problem of the Landau levels for the Dunkl-Klein-Gordon oscillator. We have solved this problem in two ways, algebraic and analytical. In the algebraic form we were able to obtain the energy spectrum of our problem by introducing a proper realization of the $su(1,1)$ Lie algebra. For the analytical form, a suitable variable change was introduced to translate the equation of our problem into a general known differential equation. Here, we obtained the energy spectrum and the radial eigenfunctions. In both cases we obtained that the solutions are consistent. Also, we have showed that all the results obtained in this paper are adequately reduced to those reported in Ref. \cite{arxiv} when the magnetic field vanishes.  Thus, we have found another physical system in which the reflection operators are involved that is exactly soluble.

Following the ideas of the present work, we have shown that the Schr\"odinger equation for the Dunkl-Landau levels has exact solutions, which is a work that will be submitted for review.

\section*{Acknowledgments}
This work was partially supported by SNI-M\'exico, COFAA-IPN, EDI-IPN, EDD-IPN, and CGPI-IPN Project Numbers $20200225$ and $20200113$.

\end{document}